\documentclass [12pt]{article}
\def\maj#1{\ifmmode\mbox{\usefont{U}{msb}{m}{n}#1}\else{\usefont{U}{msb}{m}{n}#1}\fi}
\def\v#1{\mathbf{#1}}

\begin{document}

\title{\textbf{Scattering rates and
lifetime \\ of exact and boson excitons}}
\author{M. Combescot and O. Betbeder-Matibet
 \\ \small{\textit{GPS, Universit\'e Pierre et Marie
Curie and Universit\'e Denis Diderot,
CNRS,}}\\ \small{\textit{Campus Boucicaut, 140 rue de
Lourmel, 75015 Paris, France}}}
\date{}
\maketitle

\begin{abstract}
Although excitons are not exact
bosons, they are commonly
treated as such provided that their composite
nature is included in effective
scatterings dressed by exchange. We
here \emph{prove} that, \emph{whatever these 
scatterings are}, they cannot give both the exciton
scattering rates $T_{ij}^{-1}$ and 
lifetime $\tau_0$, correctly: A striking
factor 1/2 exists between $\tau_0^{-1}$ and the sum of
$T_{ij}^{-1}$'s, which originates from the composite
nature of excitons, irretrievably lost when they
are bosonized. This result, which appears as very
disturbing at first, casts major doubts on
the overall validity of bosonization for problems
dealing with
\emph{interacting} excitons.
\end{abstract}

\vspace{2cm}

PACS.: 71.35.-y Excitons and related phenomena
			   
\hspace{1.2cm}	

\newpage

Composite particles made of two
fermions, like the excitons, are known to differ from
exact bosons. It is however accepted that they can be
treated as such, provided that their
composite nature is included through
effective scatterings 
dressed by exchange [1].

We here show that, in spite of the
wide literature on bosonization [2] which claims to
validate such a replacement, \emph{excitons can
definitely differ from bosons, in physical
effects linked to their interactions}: Indeed, the
lifetime $\tau_0$ (due to exciton-exciton
interaction) of the $N$-exciton state made with
all excitons in their ground state
0, and the
scattering rates
$1/T_{ij}$ from this state towards 
states in which two excitons out of $N$ are scattered
to
$(i,j)\neq (0,0)$, are linked by
\begin{equation}
\tau_0^{-1}=\alpha\,\sum_{(i,j)\,
couples}T_{ij}^{-1}\ ,
\end{equation}
with $\alpha=1$ if the excitons are bosonized,
and $\alpha=1/2$ if their composite nature is
kept. This result proves that it is impossible to
find effective scatterings for boson-excitons, giving
\emph{both} the lifetime and the scattering rates
correctly.

 The factor
$\alpha=1/2$ can be hard to accept at first
because we could na\"{\i}vely think that the
probability to stay in the initial
state plus the sum of probabilities
to go to all other possible states must be equal to 1.
As explained at the end of this letter, this
na\"{\i}ve thinking in fact fails for excitons, due
to a quite fundamental reason linked to
\emph{their composite nature, irretrievably lost when
they are bosonized}.

Many works have been devoted to exciton-exciton 
(X-X) interactions [3-14]. They however [15]
accept or end with an effective bosonic
Hamiltonian, so that they do not tackle the
consequences of transforming exact excitons into 
bosons. To study this problem,
we have developed \emph{a new many-body
theory} [16-18],
which allows \emph{to treat interactions between
composite bosons}, all previous many-body theories
dealing  with true fermions or
true bosons only. The case of composite excitons
is far more complex. (i) The carriers
being  indistinguishable, there is no way to know if
a given exciton is made of
$(e,h)$ or
$(e',h)$. (ii) Due to this uncertainty, there is no
way to extract from the Coulomb terms of the
semiconductor Hamiltonian
$H$, the part corresponding to a potential $V_{XX}$
between exciton operators. With $H$ not written as
$H_X+V_{XX}$, 
perturbation theory, which is
the standard way to solve problems dealing with
interactions, cannot be used anymore. (iii)  Far
worse,  excitons feel each other not only through
Coulomb interaction, but also through Pauli exclusion.
This
``Pauli interaction'', which originates from
statistical departure, can \emph{a priori} exist in
the absence of any Coulomb process. It is the
conceptually new part of our many-body theory. It
requires a very peculiar procedure to have it
appearing.

As most readers are not yet familiar with this new
theory [16-18], let us recall its main
points.  
\emph{ Two kinds of ``scatterings'' between
excitons}, instead of one, appear in this theory: (i)
$\xi_{mnij}^\mathrm{dir}$, which has the dimension of
an energy, corresponds to direct Coulomb processes
between both the ``in'' excitons
$(i,j)$ and the ``out'' excitons $(m,n)$. (ii) 
$\lambda_{mnij}$, which is dimensionless, comes from
``Pauli interaction'', without any Coulomb
contribution. In
$\xi_{mnij}^\mathrm{dir}$, the ``in'' and ``out''
excitons
are made with the same pairs, while in 
$\lambda_{mnij}$ they just \emph{exchange}
their electrons (eqs.\ (26,28) of ref.\ [16]), 
\begin{eqnarray}
\xi_{mnij}^\mathrm{dir}=\frac{1}{2}\,\int d\v r_e\,d\v
r_{e'}\,d\v r_h\,d\v r_{h'}\,\phi_m^\ast(\v r_e,\v
r_h)\,\phi_n^\ast(\v r_{e'},\v
r_{h'})\,\left[V_{ee'}+V_{hh'}-V_{eh'}-V_{e'h}\right]
\nonumber \\
\times\ \phi_i(\v r_e,\v r_h)\,\phi_j(\v r_{e'},\v
r_{h'})\ +(m\leftrightarrow n)\ ,
\end{eqnarray}
\begin{equation}
\lambda_{mnij}=\frac{1}{2}\,\int d\v r_e\,d\v r_{e'}\,
d\v r_h\,d\v r_{h'}\,\phi_m^\ast(\v r_{e'},\v r_h)\,
\phi_n^\ast(\v r_e,\v r_{h'})\,\phi_i(\v r_e,\v r_h)\,
\phi_j(\v r_{e'},\v r_{h'})\ +(m\leftrightarrow n)\ .
\end{equation}
$\phi_i(\v r_e,\v r_h)=\langle\v r_e,\v
r_h|B_i^\dag|v\rangle$ is the wave function of the $i$
exciton at hand (Wannier, Frenkel, any type).
$B_i^\dag$, which creates this $i$ exciton, is such
that 
$(H-E_i)B_i^\dag|v\rangle=0$, where $H$ is the
semiconductor Hamiltonian, made of the electron and
hole kinetic energies plus the e-e, h-h, e-h Coulomb
interactions. For bound states, we can show that
$\lambda_{mnij}\simeq \mathcal{V}_X/\mathcal{V}$ and
$\xi_{mnij}^\mathrm{dir}\simeq
R_X\mathcal{V}_X/\mathcal{V}$, with $\mathcal{V}$,
$\mathcal{V}_X$ and
$R_X$ being the sample volume, the exciton volume and
the exciton Rydberg.  

Due to
the composite nature of the excitons, we have (eq.\
(7) of ref.\ (16)),
\begin{equation}
G_{ijmn}=\langle v|B_iB_jB_m^\dag B_n^\dag|v\rangle=
\delta_{im}\delta_{jn}+\delta_{in}\delta_{jm}-2\lambda
_{ijmn}\ ,
\end{equation}
which shows that the exciton states are non-orthogonal,
while (eq.\ (5) of ref.\ (16)),
\begin{equation}
B_i^\dag\,B_j^\dag=-\sum_{mn}\lambda_{mnij}\,
B_m^\dag\,B_n^\dag\ ,
\end{equation}
which shows that the exciton states form an
overcomplete set. (Let us recall that for
boson-excitons, the Pauli scatterings $\lambda_{mnij}$
reduce to zero).

Any matrix
element between $N$-exciton states can be calculated in
terms of $\xi_{mnij}^\mathrm{dir}$
and $\lambda_{mnij}$. A major difficulty
however remains if we want to work with excitons:
While for boson-excitons, the exact Hamiltonian $H$ is
replaced by
$H_X+V_{XX}$, so that we can use the Fermi golden rule
to get the lifetime and scattering rates, there is no
such a $V_{XX}$ for exact excitons, so that we
first have to construct unconventional expressions of
these quantities in which $V$ does not appear. 

\noindent\textbf{Lifetime and
scattering rates in terms of $H$}

The time evolution of an initial state
$|\psi_0\rangle$ can be written as  
\begin{equation}
|\psi_t\rangle=e^{-i\hat{H}t}|\psi_0\rangle=|\psi_0
\rangle+|\tilde{\psi}_t\rangle\ ,
\end{equation} 
where $\hat{H}=H-\langle
H\rangle$, with $\langle
H\rangle=\langle\psi_0|H|\psi_0\rangle$. This gives
\begin{equation}
|\tilde{\psi}_t\rangle=F_t(\hat{H})\,\hat{H}|
\psi_0\rangle\ ,
\end{equation}
$F_t(E)=(e^{-iEt}-1)/E$ being such that
$|F_t(E)|^2=2\pi\, t\,\delta_t(E)$, where
$\delta_t(E)=(\pi E)^{-1}\sin(Et/2)$ is the usual
delta function of width $(2/t)$. 

The $|\psi_0\rangle$ lifetime,
defined as 
$\left|\langle
\psi_0|\psi_t\rangle\right|^2\simeq 1-t/\tau_0$, then
reads
$t/\tau_0\simeq -(\langle\psi_0|\tilde{\psi}_t\rangle
+c.c)$, which is nothing but $\langle\tilde{\psi}_t|
\tilde{\psi}_t\rangle$, due to
$\langle\psi_t|\psi_t\rangle=1$.
This leads to
\begin{equation}
t/\tau_0\simeq
\langle\psi_0|\hat{H}\,\left|F_t(\hat{H})
\right|^2\,\hat{H}\,|\psi_0\rangle
\ .
\end{equation}
As for the scattering rate from $|\psi_0\rangle$ to an
arbitrary state
$|\phi_n\rangle$, defined as $t/T_n\simeq
|\langle\phi_n|\tilde{\psi} _t\rangle|^2$ [19], it
reads, due to eq.\ (7)
\begin{equation}
t/T_n\simeq\left|\langle\phi_n|F_t(\hat{H})\,
\hat{H}\,|\psi_0\rangle\right|^2\ .
\end{equation}

These unconventional expressions of $\tau_0$ and
$T_n$ do not contain any 
$V$ as required. From them, it is however easy to
recover the conventional ones, since, for 
$H=H_0+V$ and $|\psi_0\rangle=|0\rangle$ with
$(H_0-\mathcal{E}_n)|n\rangle=0$, we have
$\hat{H}\,|\psi_0\rangle= 
\sum_{n\neq 0}|n\rangle\langle
n|V|0\rangle$. 

\noindent\textbf{Exact excitons}

We first consider the lifetime and scattering rates
of exact excitons.
Let us take as initial state, $|\psi_0\rangle=B_0
^{\dag N}|v\rangle/\sqrt{N!F_N}$, where $B_0^\dag$ is
the creation operator for one exciton in its ground
state, while
$F_N$ is such that $N!F_N=\langle
v|B_0^NB_0^{\dag N} |v\rangle$ for
$|\psi_0\rangle$ to be normalized [20]. 
This state can be seen as the one coupled to
$N$ identical photons [21] tuned on the ground
state exciton, \emph{before} its relaxation to the
$N$-pair ground state --- otherwise this initial state
would not change with time.

The reader having difficulties with
the various factors $N$ appearing in this work,
can just take $N=2$. The $N$-exciton problem
then reduces to a far simpler 2-exciton problem, 
which still has the
striking factor $\alpha=1/2$ appearing. The
various $N$'s of the final results are
actually the same for exact and boson excitons, as
physically expected: A difference in these $N$
dependences induced by the composite nature of the
exciton, would be even harder to accept than the
factor $\alpha=1/2$\,!

To get $\hat{H}\,|\psi_0\rangle$ appearing in eqs.\
(8,9), we use our new theory. We
first calculate
$\left[H,B_0^{\dag N}\right]$ by induction
from eqs.\ (14,20) of ref.\ [18], namely
$\left[H,B_i^\dag\right]=E_iB_i^\dag+V_i^\dag$
and
$\left[V_i^\dag,B_j^\dag\right]=\sum_{mn}
\xi_{mnij}^\mathrm{dir}B_m^\dag B_n^\dag$, where,
due to eq.\ (5), $\xi_{mnij}^\mathrm{dir}$
can be replaced by
$(-\xi_{mnij}^\mathrm{in})$, or better by
\begin{equation}
\hat{\xi}_{mnij}=(\xi_{mnij}^\mathrm{dir}
-\xi_{mnij}^\mathrm{in})/2=-\sum_{rs}\lambda_{mnrs}
\,\hat{\xi}_{rsij}\ .
\end{equation}
The ``in'' exchange Coulomb scattering, defined as
$\xi_{mnij}^\mathrm{in}=\sum_{rs}\lambda_{mnrs}
\xi_{rsij}^\mathrm{dir}$, 
reads as 
$\xi_{mnij}^\mathrm{dir}$ given in eq.\ (2), 
with $\phi_m^\ast(\v r_e,\v r_h)
\phi_n^\ast(\v r_{e'},\v r_{h'})$ replaced by
$\phi_m^\ast(\v r_{e'},\v r_h)
\phi_n^\ast(\v r_e,\v r_{h'})$, so that,
using eq.\ (3), we have
$\sum_{rs}\lambda_{mnrs}\xi_{rsij}^\mathrm{in}=
\xi_{mnij}^\mathrm{dir}$.
From the obtained
$\left[H,B_0^{\dag N}\right]$, we find 
\begin{equation}
\hat{H}B_0^{\dag
N}|v\rangle=(1/2)N(N-1)\,B_0^{\dag
N-2}\left[\Delta_0\,B_0^{\dag
2}+\sum_{mn\neq 00}
\hat{\xi}_{mn00}B_m^\dag
B_n^\dag \right]|v\rangle\ ,
\end{equation}
with
$\Delta_0=\hat{\xi}_{0000}+[NE_0-\langle
H\rangle]/[N(N-1)/2]$. Using $\langle
H\rangle$ calculated in ref.\ [22], this $\Delta_0$
actually reads $\hat{\xi}
_{0000}(-1+O(\eta))$, so that $\hat{H}B_0^{\dag
N}|v\rangle$ is first order in Coulomb scattering, as
physically expected.

We can then use eq.\ (5) to rewrite $\Delta_0B_0^{\dag
2}$ as a sum of $\Delta_0\lambda_{mn00}B_m^\dag
B_n^\dag$. Since these terms are one Pauli scattering
smaller than the $\hat{\xi}_{mn00}B_m^\dag B_n^\dag$
terms of the sum appearing in eq.\ (11), we
can,  at lowest order in the exciton interactions, drop
$\Delta_0B_0^{\dag 2}$ in front of the sum of eq.\
(11), so that at this order, 
$\hat{H}B_0^{\dag N}|v
\rangle$ is made of states with two
excitons $(m,n)$ outside 0. 

To get the
transition rate $T_{ij}^{-1}$ from $|\psi_0\rangle$
to one of these states $|\phi_{ij\neq 00}
\rangle= a_{ij} B_0^{\dag N-2}B_i^\dag
B_j^\dag|v\rangle$, with $a_{ij}$ being a
normalization factor, we make
$F_t(\hat{H})$ acting on the left in eq.\ (9). Since
$\hat{H}|\psi_0\rangle$ is first order in the
interactions already, we can replace
$F_t(\hat{H})$ by its free-exciton contribution, $F_t(
E_i+E_j-2E_0)$. We are left with the scalar product of
$N$-exciton states. For $N=2$, it is just $G_{ijmn}$
given in eq.\ (4), while in the large $N$ limit, and
for $(i,j,m,n)\neq 0$, we find [23]
\begin{equation}
\langle v|B_iB_jB_0^{N-2}B_0^{\dag N-2}B_m^\dag
B_n^\dag|v\rangle\simeq
G_{ijmn}\,(N-2)\,!\,F_{N-2}\ .
\end{equation}
Since, due to eqs.\ (4,10),
$\sum_{mn}G_{ijmn}\hat{\xi}_{mn00}=4\hat{\xi}_{ij00}$,
eqs.\ (9-12) lead to 
\begin{eqnarray}
T_{i\neq j}^{-1}&\simeq&t^{-1}(1/4)
N(N-1)\left|F_t(E_i+E_j-2E_0)\sum_{mn\neq 00}
G_{ijmn}\hat{\xi}_{mn00}\right|^2\nonumber
\\ &\simeq& 
2\pi\,N(N-1)\, |\xi_{ij00}
^\mathrm{dir}-\xi_{ij00}^\mathrm{in}|^2\,
\delta_t(E_i+E_j-2E_0)\ ,
\end{eqnarray}
the transition rate to $|\phi_{ii}\rangle$
being zero, due to energy and momentum
conservation. 

If we now turn to the $|\psi_0\rangle$ lifetime
(eq.\ (8)), we find, in the same way, that it reads
[24]
\begin{eqnarray}
\tau_0^{-1}&\simeq& t^{-1}(1/4)N(N-1)\sum_{0\neq
i\neq j\neq 0} |F_t(E_i+E_j-2E_0)|^2
\hat{\xi}_{00ij}\sum_{mn\neq00}G_{ijmn}\hat{\xi}_{mn00}
\nonumber \\ &\simeq& \frac{1}{4}\sum_{0\neq i\neq
j\neq 0}T_{ij}^{-1}=\frac{1}{2}\sum_{(i,j) couples}
T_{ij}^{-1}\ ,
\end{eqnarray}
as claimed in eq.\ (1).

\noindent\textbf{Boson excitons}

Bosonizing the excitons corresponds to replace the
exact exciton operators
$B_i$ by $\bar{B}_i$'s, with
$[\bar{B}_i,\bar{B}_j^\dag]=\delta_{ij}$, and the
exact Hamiltonian $H$ by 
$H_X+V_{XX}$, with
$H_X=\sum_iE_i\bar{B}_i^\dag\bar{B}_i$ and
$V_{XX}=(1/2)
\sum_{mnij}\xi_{mnij}^\mathrm{eff}\bar{B}_m^\dag
\bar{B}_n^\dag\bar{B}_i\bar{B}_j$. This leads to
replace the $G_{ijmn}$ of eq.\ (4) by
$\bar{G}_{ijmn}=\delta_{im}\delta_{jn}+\delta_{in}
\delta_{jm}$ and 
the $\hat{\xi} _{mn00}$ of eq.\ (11) by $\xi_{mn00}
^\mathrm{eff}$. In the same
way, we are led to bosonize the initial state by taking
it as
$|\bar{\psi}_0\rangle=\bar{B}_0^{\dag
N}|v\rangle/\sqrt{N!}$. We then find that the
first lines of eqs.\ (13-14) are still valid with
$\sum_{mn}\bar{G}_{ijmn}\xi_{mn00}^\mathrm{eff}$ now
equal to $2\xi_{ij00}^\mathrm{eff}$. (Note that
the similar sum for exact excitons gives 
$4\hat{\xi}_{ij00}$: This is actually the
\emph{mathematical reason for the factor 2 change}
between the
$\alpha$'s of eq.\ (1)). This leads us to find
\begin{equation}
\bar{T}_{i\neq j}^{-1}\simeq 2\pi\,
N(N-1)\,|\xi_{ij00}^\mathrm{eff}|^2\,\delta_t(E_i+E_j
-2E_0)\ ,
\end{equation}
\begin{equation}
\bar{\tau}_0^{-1}\simeq\frac{1}{2}\sum_{0\neq i\neq
j\neq 0}
\bar{T}_{ij}^{-1}=\sum_{(i,j)couples}\bar{T}_{ij}^{-1}
\ ,
\end{equation} 
which is just what
the usual Fermi golden rule readily gives. 

In order to grasp \emph{why exact and boson
excitons can have a different $\alpha$}, let us note
that the link between
$\bar{\tau}_0^{-1}$ and the
$\bar{T}_{ij}^{-1}$'s can be recovered, without
calculations, just from the closure relation
between bosons. This relation reads
\begin{eqnarray}
1 &=&(1/N!) \sum_{i_1,\cdots,i_N}\bar{B}_{i_1}
^\dag\cdots \bar{B}_{i_N}^\dag|v\rangle\langle v|
\bar{B}_{i_N}\cdots\bar{B}_{i_1}\nonumber \\
&=& |\bar{\psi}_0\rangle\langle\bar{\psi}_0|+\sum_{
i\neq 0}|\bar{\phi}_i\rangle\langle\bar{\phi}_i|
+\frac{1}{2}\sum_{(i,j)\neq 0}(1+\delta_{ij})
|\bar{\phi}_{ij}\rangle\langle\bar{\phi}_{ij}|
+\cdots\ .
\end{eqnarray} 
where  $|\bar{\phi}_i\rangle$,
$|\bar{\phi}_{ij}\rangle$, \ldots are the
\emph{normalized} boson-exciton states 
 with one, two,
\ldots boson-excitons outside 0. 
By inserting eq.\ (17) into $1=\langle\bar{\psi}_t
|\bar{\psi}_t\rangle$, we get,
to lowest order in the interactions, 
\begin{equation}
1\simeq|\langle\bar{\psi}_0|\bar{\psi}_t\rangle|^2
+(1/2)\sum_{0\neq i\neq j\neq
0}|\langle\bar{\phi}_{ij}|
\bar{\psi}_t\rangle|^2,
\end{equation}
as energy and momentum conservations impose
$\langle\bar{\phi}_i|\bar{\psi}_t\rangle=0=
\langle\bar{\phi}_{ii}|\bar{\psi}_t\rangle$ for
$i\neq 0$.
The first term of eq.\ (18) is just
$1-t/\bar{\tau}_0$, while  
$\left|\langle\bar{\phi}_{ij}|\bar{\psi}_t\rangle
\right|^2=t/\bar{T}_{ij}$, so that eq.\ (16) readily
follows from eq.\ (18).

\noindent\textbf{Discussion}

By comparing eqs.\ (13) to eq.\ (15), we
see that the transition rates of exact and boson
excitons are equal if we enforce
$\xi_{mnij}^\mathrm{eff}=\xi_
{mnij}^\mathrm{dir}-\xi_{mnij}^\mathrm{in}$. This
scattering differs from the one used up to
now [1], namely
$\tilde{\xi}_{mnij}^\mathrm{eff}=\xi_{mnij}^\mathrm{dir}
-\xi_{mnij}^\mathrm{out}$, with
$\xi_{mnij}^\mathrm{out}=(\xi_{ijmn}^\mathrm{in})^\ast$.
We have however shown [16] that, as
$\xi_{mnij}^\mathrm{out}\neq(\xi_{ijmn}^\mathrm{out})
^\ast$,
this $\tilde{\xi}_{mnij}^\mathrm{eff}$ is physically
unacceptable because the corresponding Hamiltonian is
non hermitian. The new
$\xi_{mnij}^\mathrm{eff}$ has also to be rejected
for the same reason.
. 

\emph{The most striking result of this letter} is
however not so much the difficulty in finding
physically reasonable effective scatterings giving the
exact scattering rates, but the link between the
inverse lifetime and the sum of scattering rates: The
additional factor 1/2 found for exact excitons
\emph{actually invalidates
the overall validity of bosonization},
because it is not possible to find a set of
$\xi_{mnij}^\mathrm{eff}$'s giving both, the lifetime
and the scattering rates, correctly. 

As the exact-exciton
calculations rely on a theory which is not yet
commonly used, the
reader can reasonably question the correctness of this
striking result which appears as very disturbing at
first. It is however easy to be, at least, convinced
that
\emph{different links
between
$\tau_0^{-1}$ and the sum of $T_{ij}^{-1}$'s for exact
and boson excitons are quite reasonable}: As shown
above, the link for boson-excitons comes from the
closure relation which exists between boson-exciton
states, these states forming an orthogonal basis. On
the opposite, as clear for $N=2$
already (see eqs.\ (4,5)), the basis
made of exact-exciton states is
\emph{not only non-orthogonal but, far worse,
overcomplete}.  Consequently, the closure relation for
boson excitons cannot be transposed to exact excitons,
just as the link between
$\tau_0^{-1}$ and the
$T_{ij}^{-1}$'s. Since this overcompleteness of the
exciton state set physically comes from the exciton
composite nature, the additional
factor 1/2  found between
$\tau_0^{-1}$ and the sum of 
$T_{ij}^{-1}$'s has to be seen as \emph{a non
trivial signature of the fact that excitons are deeply
made of two fermions}. Our letter shows in a
striking way that this composite nature can be crucial
for problems dealing with interacting excitons, in
spite of a widely spread belief.

\vspace{0.5cm}

\hbox to \hsize {\hfill REFERENCES
\hfill}

\noindent
(1) See for example, H. Haug and S. Schmitt-Rink,
\emph{Progr.\ Quantum.\ Electron.\ }\textbf{9}, 3
(1984).

\noindent
(2) For a review, see A. Klein and E.R.
Marshalek, \emph{Rev.\ Mod.\ Phys.}\
\textbf{63}, 375 (1991).

\noindent
(3) J.B. Stark \emph{et al.}, \emph{Phys.
Rev.} B \textbf{46}, 7919 (1992).

\noindent
(4) K. Bott \emph{et al.}, \emph{Phys. Rev.} B
\textbf{48}, 17418 (1993).

\noindent
(5) W. Sch\"{a}fer \emph{et al.}, \emph{Phys. Rev.} B
\textbf{53}, 16429 (1996).

\noindent
(6) J. Fern\'{a}ndez-Rossier \emph{et al.},
\emph{Phys. Rev.} B
\textbf{54}, 11582 (1996).

\noindent
(7) P. Le Jeune \emph{et al.},
\emph{Phys. Rev.} B
\textbf{58}, 4853 (1998).

\noindent
(8) H. Suzuura \emph{et al.},
\emph{Solid State Com.}\ \textbf{108}, 289 (1998).

\noindent
(9) C. Ciuti \emph{et al.}, \emph{Phys. Rev.} B
\textbf{58}, 7926 (1998).

\noindent
(10) J. Inoue \emph{et al.}, \emph{Jour.
Phys. Soc. Japan} \textbf{67}, 3384 (1998).

\noindent
(11) I. Imamoglu, \emph{Phys.\ Rev.\ Rap.\ Com}.\ B
\textbf{57}, 4195 (1998).

\noindent
(12) J. Inoue \emph{et al.}, \emph{Phys.
Rev.} B \textbf{61}, 2863 (2000).

\noindent
(13) S. Ben-Tabou de-Leon and B. Laikhtman,
\emph{Phys. Rev.} B \textbf{63}, 125306 (2001).

\noindent
(14) S. Okumura and T. Ogawa, \emph{Phys.\ Rev}.\ B
\textbf{65}, 35105 (2001).

\noindent
(15) Exact excitons are kept in ref.\ (9).
However, the procedure used is conceptually
incorrect, as easy to show from a correct
algebra (see ref.\ (18)).

\noindent
(16) M. Combescot and O. Betbeder-Matibet,
\emph{Europhys.\ Lett.}\ \textbf{58}, 87
(2002). In our previous works, $\xi^\mathrm{in}$ and
$\xi^\mathrm{out}$ were called $\xi^\mathrm{left}$ and
$\xi^\mathrm{right}$.

\noindent
(17) M. Combescot and O. Betbeder-Matibet,
\emph{Europhys.\ Lett.}\ \textbf{59}, 579
(2002).

\noindent
(18) O. Betbeder-Matibet and M. Combescot,
\emph{Eur\ Phys.\ J. B} \textbf{27}, 505
(2002).

\noindent
(19) and not as $|\langle\phi_n|\psi_t\rangle|^2$, in
order to have a quantity linear in $t$, even for
$\langle\phi_n|\psi_0\rangle\neq 0$.

\noindent
(20) While $F_N=1$ for boson excitons, $F_N\propto\exp
(-N\eta)$ for exact excitons, with $F_{N-p}/F_N\simeq
1+O(\eta)$, as shown in  
M. Combescot and C. Tanguy,
\emph{Europhys.\ Lett.}\ \textbf{55},
390 (2001), 
M. Combescot, X. Leyronas and C.
Tanguy, \emph{Eur.\ Phys.\ J.\ B} \textbf{31}, 17
(2003).

\noindent
(21) We do not consider polarization effects.
This allows to drop the spin part of the excitons,
which simplifies the notations, without
any incidence on the exact/boson exciton problem. The
results given here actually correspond to an initial
state prepared by $\sigma_+$ photons in a quantum well.

\noindent
(22) O. Betbeder-Matibet and M. Combescot,
\emph{Eur.\ Phys.\ J.\ B} \textbf{31}, 517 (2003).

\noindent
(23) The precise derivation of this $N$-exciton result
will be given elsewhere. The reader questioning this
result can stay with 
$N=2$.

\noindent
(24) This $\tau_0$ corresponds to processes in which
\emph{two} excitons, out of $N$, go to $(i,j)\neq 0$.
It is
$N$ times smaller than the lifetime which enters the
exciton linewidth and corresponds to have \emph{one}
exciton, out of $N$, going from 0 to $i\neq 0$.

\end{document}